\documentstyle[aps,preprint,prd,epsfig]{revtex}
\newcommand{\spur}[1]{\not\! #1 \,}   
\begin{document}
\title{\small \phantom{.}\hfill BARI-TH 350/99 \\[1cm]
\large
{\bf DYNAMICAL  SYMMETRY BREAKING IN PLANAR QED}}
\author{Paolo Cea$^{a,b}$ and Luigi Tedesco$^{b}$} 
\maketitle
\begin{center}
$^{a}$Dipartimento di Fisica and $^{b}$INFN, Sezione di Bari \\
Via Amendola 173,  I-70126 Bari, Italy*
\end{center}
\vspace{2cm}
\begin{abstract}
We investigate $(2+1)$-dimensional {\it QED} coupled with Dirac fermions
both at zero and finite temperature. We discuss in details two-components
(P-odd) and four-components (P-even) fermion fields. 
We focus on P-odd and P-even Dirac fermions in presence of an external 
constant magnetic field. In the one-loop approximation we find that the 
spontaneous generation of an uniform magnetic condensate  
is energetically favoured in the case of the P-odd massive theory.
Moreover, we find that the 
spontaneous generation of the magnetic condensate survives
even at infinite temperature. We also discuss the spontaneous generation
of fermion mass in presence of an external magnetic field.
\end{abstract}
\newpage

{\bf 1. Introduction}
\vskip 0.7truecm
Three dimensional  gauge theories have attracted interest since long time due to 
 the possibility of a gauge invariant mass term for the gauge 
fields~\cite{D-J-T}. Three dimensional field theoretical models are
believed to provide effective theories for long distance planar condensed
matter models~\cite{Fradkin} and high temperature four dimensional gauge field 
theories~\cite{Jackiw}. In particular quantum electrodynamics in $(2+1)$-dimensions
has been intensively studied  within the effective theories of 
high-$T_c$ superconductivity~\cite{Aitchson}. \\ 
As a matter of fact, it turns out that in $QED_{2+1}$ with massive Dirac
fermions the perturbative ground state is unstable towards the spontaneous
generation of an uniform magnetic condensate~\cite{CEA1,Hosotani}. 
Moreover, it turns out that this remarkable property is not shared with 
other $(2+1)$-dimensional field theories~\cite{CEA4}.
\\
 It is well known that four dimensional Dirac fermions with  Yukawa
coupling to a scalar field develop zero mode solutions near a domain 
wall~\cite{Rebbi}. 
Indeed, let us consider a four dimensional massless Dirac fermion coupled
to a scalar field through the Yukawa coupling :
\begin{equation}
\label{eq1.1}
{\cal {L}}_{\psi \phi} = {\cal{L}}_{\psi} + {\cal{L}}_{\phi} \;  ,
\end{equation}
where
\begin{equation}
\label{eq1.2}
{\cal{L}}_{\psi} = \bar{\psi}(x)  i \spur \partial \psi(x) - g_{Y} \bar{\psi}(x) 
\psi(x) \phi(x)  \; ,
\end{equation}
\begin{equation}
\label{eq1.3}
{\cal{L}}_{\phi}= \frac {1} {2} ( \partial_{\mu} \phi)^2 - \frac {\lambda} {4}
(\phi^2 - v^2)^2 \; .
\end{equation}
The scalar field develops a non vanishing vacuum expectation  value 
 $\langle \phi \rangle = \pm v$ . In general one may assume that
 there are regions with $\langle \phi \rangle = + v$ and 
$\langle \phi \rangle = - v$. It is easy to see that the classical equation of 
motion of the scalar field admits the solution describing the transition layer 
between two adjacent regions 
with different values of $\langle \phi \rangle$:
\begin{equation}
\label{eq1.4}
\phi(z)= v \tanh \left(\frac {z} {\Delta}\right),
\end{equation}
where
\begin {equation}
\label{eq1.5}
\Delta= \frac {1} {v \sqrt{\frac {\lambda} {2}}} 
\end{equation}
is the thickness of the wall. \\
The solution Eq.~(\ref{eq1.4}) is known as kink.
In presence of the kink the Dirac equation
\begin{equation}
\label{eq1.6}
[i \spur \partial - g_Y \phi(x)] \psi(x)=0
\end{equation}
admits zero energy solutions localized on the wall~\cite{Rebbi}.
Using the 
following representation of the Dirac algebra:
\begin{equation}
\label{eq1.7}
{\gamma}^3= i \left(
\begin{array} {cc}
0 & 1 \\
1 & 0
\end{array}
\right), \; \;
{\gamma}^{\alpha}=  \left(
\begin{array} {cc}
{\tilde{\gamma}}^{\alpha} & 0 \\
0  & -{\tilde{\gamma}}^{\alpha}
\end{array}
\right), \; \; 
\alpha = 0, 1, 2
\end{equation}
with
\begin{equation}
\label{eq1.8}
{\tilde{\gamma}}^0=\sigma_3, \; \; 
{\tilde{\gamma}}^1=i \sigma_1, \; \; {\tilde{\gamma}}^2=i \sigma_2 \; ,
\end{equation}
we find for the zero modes:
\begin{equation}
\label{eq1.9}
\psi_0(x,y,z) = \frac {N} {\sqrt{2}} \; \omega(z) \left(
\begin{array} {cc}
\rho(x,y) \\
\rho(x,y)
\end{array}
\right)
\end{equation}
\begin{equation}
\label{eq1.10}
\omega(z)= exp\left\{- g_Y \int_0^z \phi(z') d z' \right\}
\end{equation}
where  $\rho(x,y)$ is a Pauli spinor, and 
\begin{equation}
\label{eq1.11}
N=\left[\int^{+\infty}_{-\infty} d z \; \omega^2 (z) \right]^{- \frac {1} {2}}=
\sqrt{\pi} \Delta \frac {\Gamma (g_Y \Delta)} {\Gamma (g_Y \Delta+ \frac {1} 
{2})} \; .
\end{equation}
It is easy to see that $\rho(x,y)$ satisfies the ($2+1$)-dimensional massless 
Dirac equation with the Dirac algebra Eq.~(\ref{eq1.8}). \\
Thus we see that two-component Dirac fermions in $(2+1)$ dimensions are relevant
for the dynamics of zero modes localized at the domain walls. Indeed,
recently~\cite{CEA5} it has been suggested that the remarkable phenomenon
of the spontaneous generation of uniform magnetic condensate in $(2+1)$-dimensional
{\it QED} gives rise to ferromagnetic domain walls at the cosmological electroweak
phase transition which could account for the cosmological primordial magnetic
field. \\
The aim of this paper is to study in details the $(2+1)$-dimensional  
$QED$ with  two-component (parity-violating) and four-component 
(parity-invariant) fermions both at zero and finite temperature.
In particular we compare in details the $(2+1)$-dimensional $QED$
coupled with both two and four components Dirac fermions in the one-loop 
approximations. For pedagogical reasons, we present our results
in greater details with complete calculations which are not available in our 
previuos published papers.
The plan of the paper is as follows. In Sect.~2 we consider  Dirac
fermions with P-odd mass term (two component fermions) at zero and finite 
temperature.
Section~3 is devoted to the study of P-even Dirac fermions. In Sect.~4 we
discuss the spontaneous generation of a constant fermion mass triggered by
an external magnetic field. Finally our conclusions are drawn in Sect.~5.
%
\\
\\
\\
{\bf 2. P-Odd Fermions}
\\
In this section we discuss  relativistic fermions coupled with the 
electromagnetic field in ($2+1$)-dimensions. The relevant Lagrangian is: 
\begin{equation}
\label{eq2.1}
{\cal {L}}_{QED} = - \frac {1} {4} F_{\mu \nu}(x) F^{\mu \nu} (x) + \bar {\psi} (x)
[i \spur \partial - e \spur A  (x) - m ] \psi(x) \; .
\end{equation}
From the Lagrangian Eq.~(\ref{eq2.1}) it is easy to obtain the equations of motion.
In particular the fermion fields satisfy the familiar Dirac equation :
\begin{equation}
\label{eq1}
(i \; \gamma^{\mu} \; \partial_{\mu} - e \; 
\gamma^{\mu} \; A_{\mu} - m) \; \psi=0 \; .
\end{equation}
The gamma matrices satisfy the Clifford algebra :
\begin{equation}
\label{eq2}
\{\gamma^{\mu}, \gamma^{\nu}\} =2 \; \; g^{\mu \nu}, \;\;  \gamma^{\mu} 
\gamma^{\nu}= g^{\mu \nu} -i \epsilon^{\mu \nu \rho} \gamma_{\rho}, \;\;  
\gamma_{\mu}=g_{\mu \nu} \gamma^{\nu}
\end{equation}
with the flat Minkowski metric is $g^{\mu \nu} =diag(1,-1,-1)$.
\\
A spinorial representation in three dimensions is provided by two-component 
Dirac spinors. The fundamental representation of the Clifford algebra is given 
by $2 \times 2$ matrices which can be constructed from the Pauli matrices:

\begin{equation}
\label{eq3}
\gamma^0=\sigma_3, \; \; \gamma^1 =i \sigma_1, \; \; \gamma^2=i \sigma_2 \; .
\end{equation}
We can define the parity and time-reversal transformation~\cite{D-J-T}

\begin{eqnarray}
\label{eq4}
{\cal {P}}  A^0 (\vec{x}, t) {\cal {P}}^{-1} & = & A^0(\vec{x'}, t) 
\nonumber  \\
{\cal {P}}  A^1 (\vec{x}, t) {\cal {P}}^{-1} & =& - A^1(\vec{x'}, t) \\
 {\cal {P}}  A^2 (\vec{x}, t) {\cal {P}}^{-1} & =&  A^2(\vec{x'}, t)
\nonumber \\
 {\cal {P}} \; \psi(\vec{x}, t) \; 
{\cal {P}}^{-1} &=& \sigma_1 \psi(\vec{x'}, t)
\nonumber 
\end{eqnarray}
with $\vec{x}=(x_1,x_2)$ and $\vec{x'}=(-x_1,x_2)$,
\begin{eqnarray}
\label{eq5}
{\cal {T}}  A^0 (\vec{x}, t) {\cal {T}}^{-1} & = & A^0(\vec{x}, -t) 
\nonumber  \\
{\cal {T}} \; {\vec {A}} \; (\vec{x}, t) {\cal {T}}^{-1} & =&  
-{\vec {A}} (\vec{x}, -t) \\
{\cal {T}} \; \psi(\vec{x}, t) \; 
{\cal {T}}^{-1} &=& \sigma_2 \psi(\vec{x}, -t) \; .
\nonumber 
\end{eqnarray}
Note that in this representation the mass term $m {\bar {\psi}} \psi$ is 
odd under both ${\cal {P}}$
and ${\cal {T}}$ transformations~\cite{D-J-T}. It is important to point out 
that this parity-violating mass is, in fact, the only possibility in the 
two-component formalism.
\\
Let us consider, now,  the Hamiltonian in the fixed-time temporal gauge  $ A_0=0$ :
\begin{eqnarray}
\label{eq6a}
H &=& \int d^2 x \left\{  \frac {1} {2}{\vec{E}}^2(x) +\right. 
       \frac {1} {2}  B^2(x) 
      + \psi^{\dag}(x)[- i \vec{\alpha} \cdot \vec{\nabla}
      + \beta m] \psi(x)+ \nonumber \\ 
  && ~~~~~~~~~~~-  \left. e \psi^{\dag}(x) \vec{\alpha} 
     \cdot \vec{A}(x) \psi(x) \right\} \; ,
\end{eqnarray}
where $\beta= \gamma^0$ and $\vec {\alpha} = \gamma^0 \vec {\gamma}$. \\
Note that in two spatial dimensions the magnetic field is a (pseudo-)scalar.
We are interested in the case of spinorial quantum electrodynamics in presence
of a background field. So that we write :
\begin{equation}
\label{eq6c}
A_k(x)={\bar {A}}_k(x) + {\eta}_k(x).
\end{equation}
where $\eta(x)$ is the fluctuation over the background ${\bar {A}}(x)$ .
\\
We consider a background field which corresponds to an uniform magnetic field. Thus in
the so-called Landau gauge we have :
\begin{equation}
\label{eq6}
{\bar {A}}_k(x) = {\delta}_{k2} x_1 B \;\;\;\; k= 1,2 \; .
\end{equation}
Inserting Eq.~(\ref{eq6c}) into Eq.~(\ref{eq6a}), in the one loop approximation 
we rewrite the Hamiltonian  as :
\begin{eqnarray}
\label{eq6b}
H_0 &=& H_{\eta}+H_D + V \frac {B^2} {2} 
 = \int  d^2 x \left \{ \frac {1} {2} {\vec{E}}^2(x) + \frac {1} {2} 
[\epsilon_{i j} \partial_i \eta_j(x)]^2 \right\}+ \nonumber \\ 
 &&  + \int d^2 x \left \{ 
\psi^{\dag}(x)[{\alpha}_k(-i\partial_k-e\bar{A}_k)
+\beta m ] \psi(x) \right \} + V \frac {B^2} {2} 
\end{eqnarray}
where $V$ is the spatial volume. In this approximation the Dirac equation reduces to:
\begin{equation}
\label{eq7}
i \frac {\partial \psi (\vec{x}, t)} {\partial t} = [ {\vec {\alpha}} \cdot 
(-i \vec {\nabla} - e \vec{\bar {A}}) + \beta m] \; \psi(\vec{x}, t) \; .
\end{equation}
As it is well known the Dirac equation  Eq.~(\ref{eq7})
can be exactly solved~\cite{AKHEIZER}. Indeed, taking 
$$
\psi(\vec {x}, t) = e^{-iEt} \psi (\vec {x}) ,
$$
we get :
\begin{equation}
\label{eq8}
[\vec {\alpha} \cdot (-i \vec {\nabla} - e \vec {\bar {A}} ) + \beta m] \psi 
(\vec {x}) = E \psi (\vec {x}).
\end{equation}
It is straightforward to obtain the solutions of Eq.~(\ref{eq8}) (we assume $eB >0$) :
\begin{mathletters}
\begin{equation}
E= +E_n \, , \; n \geq 1 \; \; \psi^{(+)}_{n,p} = \sqrt{\frac {E_n+m} {2 E_n}} 
\frac {e^{i p x_2}} {\sqrt {2 \pi}} e^{-\frac {1} {2} {\zeta}^2}
 \left( \begin{array}{c} 
N_n H_n(\zeta) \\
- \frac {\sqrt{E_n^2-m^2}} {E_n+m} N_{n-1} H_{n-1}(\zeta)
\end{array} \right) 
\end{equation}
\begin{equation}
\label{eq9}
E= -E_n \, , \; n \geq 1 \; \; \psi^{(-)}_{n,p} = \sqrt{\frac {E_n-m} {2 E_n}} 
\frac {e^{i p x_2}} {\sqrt {2 \pi}} e^{-\frac {1} {2} {\zeta}^2}
 \left( \begin{array}{c} 
N_n H_n(\zeta) \\
- \frac {\sqrt{E_n^2-m^2}} {E_n-m} N_{n-1} H_{n-1}(\zeta)
\end{array} \right) 
\end{equation}
\begin{equation}
E=E_0=m \;\;\;\;\;\; \psi_{0,p}=N_0 \frac {e^{i p x_2}} 
{\sqrt{2 \pi}} e^{- \frac {1} 
{2} {\zeta}^2}  \left( \begin{array}{c} 1 \\0 \end{array} \right),
\end{equation}
\end{mathletters}
where the energy spectrum is :
\begin{equation}
\label{eq10}
E_n=\sqrt{ 2neB +m^2} \; 
\end{equation}
and
\begin{equation}
\label{eq11}
N_n= \frac 
{\left( \frac {eB} {\pi} \right)^{\frac {1} {4}}} 
{\sqrt{2^n n!}}, \;\; \zeta=\sqrt{eB} \left( x_1 - \frac {p} {eB} \right) \; .
\end{equation}
$H_n(x)$ is the $n$-th Hermite polynomial, $\psi^{(+)}_{n,p}$ and $\psi^{(-)
}_{n,p}$ are respectively positive and negative energy solutions which  are 
normalized as:
\begin{equation}
\label{eq12}
\int^{+\infty}_{-\infty} d^2 x \; {\psi^{(\pm)}_{n,p}}^{\dag} 
(\vec {x}) \; \psi^{(\pm)}_{n',p'} (\vec {x})
= \delta(p-p') \delta_{n n'} \; .
\end{equation}
Note that due to the zero modes Eq.~(24c) the spectrum is not charge symmetric 
(see Fig.~1). The Landau levels are infinitely degenerate with density of states:
\begin{equation}
\label{eq14}
\int^{+\infty}_{-\infty} dp \; {\psi^{(\pm)}_{n,p}}^{\dag} 
(\vec {x}) \; \psi^{(\pm)}_{n,p} (\vec {x}) = \frac {eB} {2 \pi} \; . 
\end{equation}
It is convenient to  expand the fermion field operator $\psi(\vec{x})$ in terms 
of the wave function basis $\psi^{(+)}_{n,p}$ and  $\psi^{(-)}_{n,p}$. In
other words, we adopt the so-called Furry's representation. We have:
\begin{equation}
\label{eq15}
m<0 \; \;\;\;
\psi(x)= \sum^{\infty}_{n=1} \int^{+\infty}_{-\infty} 
d p \; a_{np} \psi^{(+)}_{np} (x) + 
\sum^{\infty}_{n=0} \int^{+\infty}_{-\infty} d p \; b^{\dag}_{np} \psi^{(-)}_{np} \; ,
\end{equation}
\begin{equation}
\label{eq16}
m>0 \; \;\;\;
\psi(x)= \sum^{\infty}_{n=0} \int^{+\infty}_{-\infty}
d p \; a_{np} \psi^{(+)}_{np} (x) + 
\sum^{\infty}_{n=1} \int^{+\infty}_{-\infty}
d p \; b^{\dag}_{np} \psi^{(-)}_{np} \; .
\end{equation}
We observe that in the case  of positive mass the positive solutions have 
eigenvalues $+E_n$ with $n \geq 0$ and negative ones $-E_n$ with $n \geq 1$ and  
vice versa in the case of negative mass. In the expansion of $\psi$ we 
associate particle operators to positive energy solutions and antiparticle 
operators to negative  energy solutions. These operators satisfy the standard 
anticommutation relations:
\begin{equation}
\label{eq17}
\{a_i(n,p),a^{\dag}_j(n',p')\}=\{b_i(n,p),b^{\dag}_j(n',p')\}=\delta_{ij} 
\; \delta_{n n'} \; \delta(p-p')
\end{equation}
the others anticommutators being zero.
The Dirac Hamiltonian operator

\begin{equation}
\label{eq17bis}
H_D= \int d^2  x \; [ \psi^{\dag} (x) ( -i \vec {\alpha} \cdot \vec{\nabla} -
{\vec {\alpha}} \cdot {\vec {\bar {A}}} +
\beta m ) \; \psi (x)]
\end{equation}
in the Furry's representation reads:
\begin{equation}
\label{eq18}
m>0\;\;\; H_D=\int_{-\infty}^{+\infty} d\;p \left[ \sum_{n=0}^{\infty} E_n 
a^{\dag}_{np} a_{np} +\sum_{n=1}^{\infty} E_n b^{\dag}_{np} b_{np} \right] -
\frac {eB} {2\pi} V \sum_{n=1}^{\infty} E_n
\end{equation}
\begin{equation}
\label{eq19}
m<0\;\;\; H_D=\int_{-\infty}^{+\infty} d\;p \left[ \sum_{n=1}^{\infty} E_n 
a^{\dag}_{np} a_{np} +\sum_{n=0}^{\infty} E_n b^{\dag}_{np} b_{np} \right] -
\frac {eB} {2\pi} V \sum_{n=0}^{\infty} E_n \; .
\end{equation}
Let us, firstly, evaluate the fermion condensate at zero temperature. By
using the expansions 
Eqs.~(\ref{eq15}) and (\ref{eq16}), it is straightforward to obtain:
\begin{equation}
\label{eq20}
m>0 \;\;\;\; \langle 0 | {\bar {\psi}} \psi | 0  \rangle 
= - \frac {m eB} {2 \pi} 
\sum_{n=1}^{\infty} \frac {1} {E_n}
\end{equation}
and
\begin{equation}
\label{eq21}
m<0 \;\;\;\; \langle 0 | {\bar {\psi}} \psi | 0 \rangle 
= - \frac {m eB} {2 \pi} 
\sum_{n=0}^{\infty} \frac {1} {E_n} \; .
\end{equation}
In order to evaluate the massless limit, we observe that
\begin{equation}
\label{eq22}
\sum_{n=1}^{\infty} \frac {1} {E_n} = \frac {1} {\sqrt{ 2 e B}} 
\sum_{n=1}^{\infty} \frac {1} {\sqrt{n+x}}
\end{equation}
where $x= \frac {m^2} {2 e B}$. We recognize in Eq.~(\ref{eq22}) the Hurwitz zeta 
function that in the massless limit ($x \rightarrow 0$) reduces to the 
Riemann zeta function $\zeta(\frac {1} {2}) \simeq -1.46035451$. 
So that in the massless limit we get~\cite{CEA4}:
\begin{equation}
\label{eq23}
\lim_{m \rightarrow 0^-} \langle 0|{\bar {\psi}} 
\psi | 0 \rangle = \frac {eB} {2 
\pi}
\end{equation}
and
\begin{equation}
\label{eq24}
\lim_{m \rightarrow 0^+} \langle 0 | {\bar {\psi}} \psi | 0 \rangle = 0 \; .
\end{equation}
Therefore we see that in the massless limit the magnetic field gives origin to a fermion 
condensate only in the case of a vanishing negative mass term.
\\
We consider, now, the background field effective action. As it is well known, in
the one-loop approximation the background field effective action coincides with
the vacuum energy in presence of the static background field. In our case we need
to evaluate the vacuum energy density in presence of the constant magnetic field
$B$. \\
To this end we need to evaluate the Dirac energy, $E_D$, defined as : 
\begin{equation}
\label{eq25}
E_D= \frac {\langle 0 | H_D | 0 \rangle} { \langle 0 | 0 \rangle},
\end{equation}
where $| 0 \rangle$ is the fermion vacuum.
By using  Eqs.~(\ref{eq18}) , (\ref{eq19}), it is straightforward to 
obtain the Dirac energy:
\begin{equation}
\label{eq27}
m<0 \;\;\; \;  E_D(B)= - \frac {eB} {2 \pi} V \sum_{n=0}^{\infty} E_n
\end{equation}
and
\begin{equation}
\label{eq28}
m>0 \;\;\; \;  E_D(B)= - \frac {eB} {2 \pi} V \sum_{n=1}^{\infty} E_n \; .
\end{equation}
 Note that  Eqs.~(\ref{eq27})
and (\ref{eq28})  differ in the term $|m| \frac {eB} {2 \pi} V$. 
\\
To perform the sums in Eqs.~(\ref{eq27}) and (\ref{eq28}) we use the integral 
representation :
\begin{equation}
\label{eq29}
\sqrt{a} = - \int_0^{\infty} \frac {d s} {\sqrt{ \pi s}} \; \frac {d} {d s} \; 
 e^{- a s}  \; .
\end{equation}
We have :
\begin{equation}
\label{eq29bis}
 - \frac {eB} {2 \pi} V \sum_{n=0}^{\infty} E_n \,
= \frac {eB} {2 \pi} V \int_0^{\infty} \frac {ds} {\sqrt{\pi s}} \frac 
{d} {ds} \left( \frac {e^{-m^2 s}} { 1- e^{-2eBs}} \right) \; . 
\end{equation}
Let us consider the negative mass case. By using Eq.~(\ref{eq29bis}) we rewrite
 Eq. (\ref{eq27}) as follows :
\begin{equation}
\label{eq30}
E_D(B)= \frac {eB} {2 \pi} V \int_0^{\infty} \frac {ds} {\sqrt{\pi s}} \frac 
{d} {ds} \left( \frac {e^{-m^2 s}} { 1- e^{-2eBs}} \right) \; , 
\end{equation}
whence, after subtracting $E_D(B=0)$,
\begin{equation}
\label{eq31}
E_D(B)-E_D(B=0) = \frac {eB} {2 \pi} V \int_0^{\infty} \frac {ds} {\sqrt{\pi 
s}} \frac {d} { ds} \left( \frac {e^{-m^2s}} {1-e^{-2eBs}} - 
\frac {e^{-m^2 s}} {2eBs}\right) \; .
\end{equation}
Introducing the dimensionless variable  $\lambda= \frac {eB} {m^2}$ and the function
\begin{equation}
\label{eq32}
g(\lambda) =\int_0^{\infty} \frac {ds} 
{\sqrt{\pi s}} \; \frac {d} {d s} \; 
\left[ \frac {e^{- \frac {s} {\lambda}}} {1-e^{-2s}}
- \frac {e^{- \frac {s} {\lambda}}} {2s} \right],
\end{equation}
we obtain the vacuum energy density ${\cal {E}} =\frac {E} {V}$:
\begin{equation}
\label{eq33}
{\tilde {\cal {E}}}_{m<0} = {\cal {E}}_{m<0}(B) - {\cal {E}}(0)= \frac {B^2} 
{2} + \frac {(eB)^{\frac {3} {2}}} {2 \pi} g\left(\frac {eB} {m^2}\right) \, ,
\end{equation}
where we added the classical magnetic energy density. \\
The case of positive mass can be handled in the same way. We get :
\begin{equation}
\label{eq34}
{\tilde {\cal {E}}}_{m>0} = {\cal {E}}_{m>0}(B) - {\cal {E}}(0)= \frac {B^2} 
{2} + \frac {(eB)^{\frac {3} {2}}} {2 \pi} 
g\left(\frac {eB} {m^2}\right) + \frac {eB} {2 
\pi} |m| \; .
\end{equation}
We introduce the further dimensionless parameter $\alpha=\frac {|m|} {e^2}$ to
 rewrite Eqs.~(\ref{eq33}) and (\ref{eq34}) as :
\begin{equation}
\label{eq35}
\frac {\tilde {\cal{E}}_{m<0}} {|m|^3} = \frac {{\lambda}^2 \alpha} {2} + 
\frac {\lambda^{\frac {3} {2} }} {2 \pi} g(\lambda) \; ,
\end{equation}
\begin{equation}
\label{eq36}
\frac {\tilde {\cal{E}}_{m>0}} {|m|^3} = \frac {{\lambda}^2 \alpha} {2} + 
\frac {\lambda^{\frac {3} {2} }} {2 \pi} g(\lambda) + \frac {\lambda} {2 \pi} \; .
\end{equation}
In Figures~2 and  3 we plot  Eq.~(\ref{eq35}) and (\ref{eq36}) respectively as 
a function of $\lambda$ for two different values of  $\alpha$. \\
As it is evident the background field effective potential displays a negative
minimum only in the case of fermions with negative 
mass term~\cite{CEA1}. Indeed,  using the expansion : 
\begin{equation}
\label{eq37}
g(\lambda)  \stackrel{\lambda \rightarrow  0}{\sim} 
- \frac {1} {2 \lambda^{\frac {1} {2}}} +
\frac {\lambda^{1\over 2}} {12} \; ,
\end{equation}
we have : 
\begin{equation}
\label{eq38}
{\tilde {\cal {E}}}_{m<0}
 \sim - \frac {eB} {4 \pi} |m| + \frac {1} {24 \pi} \frac {(eB)^2}
{|m|} + \frac {B^2} {2}.
\end{equation}
We see that  the negative linear term, which is absent in the positive mass case,
 is responsible of the negative minimum in vacuum energy density.
\\
 It is worthwhile to evaluate the thermal 
corrections to the condensate and to the vacuum energy density. Our approximation
corresponds to perform the thermal average with respect to the one-loop Hamiltonian
$H_0$, Eqs.~(\ref{eq6b}). We introduce the symbol 
$\langle ... \rangle_{\beta}$  to mean the thermal average with respect to the 
Hamiltonian $H_0$.
Let us observe that 
$\langle a^{\dag}_{np} a_{np} \rangle_{\beta}$ and 
$\langle b^{\dag}_{np} b_{np} \rangle_{\beta}$ can be expressed in
terms of the equilibrium occupation number $n(p)$:
\begin{equation}
\label{eq39}
n(p)=
\langle a^{\dag}_{np} a_{np}\rangle_{\beta}
=1- \langle a_{np} a^{\dag}_{np}\rangle_{\beta}=
\frac {1} { e^{\beta E_n}+1} \; , 
\end{equation}
\begin{equation}
\label{eq40}
1-n(p)=
\langle b_{np} b_{np}^{\dag}\rangle_{\beta}
=1- \langle b^{\dag}_{np} b_{np}\rangle_{\beta} =
\frac {1} { e^{-\beta E_n}+1} \; .
\end{equation}
It is, now, easy to evaluate the fermion condensate at finite temperature.
 We have for $m<0$:
\begin{equation}
\label{eq40bis}
\langle {\bar {\psi}} \psi {\rangle}_{\beta}= 
\sum_{n=1}^{\infty} \int dp \;  
\langle {a^{\dag}}_{np} a_{np} {\rangle}_{\beta} 
{{\psi}^{(+)}_{n,p}}^{\dag} \gamma^0 
{\psi}^{(+)}_{n,p} + 
\sum_{n=0}^{\infty} \int dp \;
\langle  b_{np} {b^{\dag}}_{np}  {\rangle}_{\beta} 
{{\psi}^{(-)}_{n,p}}^{\dag} \gamma^0 
{\psi}^{(-)}_{n,p} \; ,
\end{equation}
with ${\psi}^{(+)}_{n,p}$ and ${\psi}^{(-)}_{n,p}$ given by Eqs.~(24a,b,c).
\\
Using Eq.~(\ref{eq39}) and Eq.~(\ref{eq40}) we obtain  
 the fermion condensate at finite 
temperature:
\begin{equation}
\label{eq41}
\langle {\bar {\psi}} \psi {\rangle}_{\beta} = \frac {eB} {2 \pi} 
\frac {1} {e^{\beta m}+1} - 
\frac  {m e B} { 2 \pi} \sum_{n=1}^{\infty} \frac {1} 
{E_n} \tanh \left( \frac {\beta} {2} E_n\right).
\end{equation}
One can check that this last expression holds for both positive and negative mass. \\
If we perform the massless limit, then we face with the problem of the order of the limits
 $m \rightarrow 0$ and $\beta \rightarrow \infty$. Indeed  we have:
\begin{equation}
\label{eq42}
    \lim_{m \rightarrow 0^+} \; \lim_{\beta \rightarrow \infty}
 \langle {\bar {\psi}} \psi {\rangle}_{\beta} = 0 \;\;\; 
    \lim_{m \rightarrow 0^-} \; \lim_{\beta \rightarrow \infty}
 \langle {\bar {\psi}} \psi {\rangle}_{\beta}= \frac {eB} {2 \pi} \; ,
\end{equation}
while
\begin{equation}
\label{eq43}
\lim_{\beta \rightarrow \infty} \;
\lim_{m \rightarrow 0^-} \langle {\bar {\psi}} \psi 
\rangle_{\beta} =
\lim_{\beta \rightarrow \infty} \;
\lim_{m \rightarrow 0^+} \langle \bar {\psi} \psi 
\rangle_{\beta} = \frac {eB} {4 \pi} \; .
\end{equation}
Of course Eq.~(\ref{eq42}) agrees with the zero temperature result. On the other hand,
Eq.~(\ref{eq43}) shows that the massless limit of the fermion condensate 
at any temperature is T-independent and symmetric in $m$ . 
 Thus, the fermion condensate 
$\langle \bar {\psi} \psi \rangle_{\beta}$ is a non-analytic function at the 
origin in the $(m,T)$-plane and its limit value depends on the order of the zero-mass  
and $\beta$-infinity limit . This non-analitycity of the thermal 
condensate  is also present in the P-invariant formulation of the theory~\cite{D-H},
 as we will see in the next Section. A similar  non-analyticity
 has been noticed by the authors of Ref.~\cite{N-S}  in the 
study of induced quantum numbers at finite temperature. 
It is worthwhile to stress that the presence of a heat bath, even with an infinitesimal 
temperature, makes the massless limit of the condensate symmetric and different from zero. 
Moreover, from Eq.~(\ref{eq43}) we see that the fermion condensate of the zero
mass theory in presence of an external constant magnetic field  corresponds to 
half filled zero modes~\cite{CEA3}.  \\
Let us turn on  the finite temperature effective potential. As it is well known, the 
relevant quantity is the free energy density. We have :
\begin{equation}
\label{eq44}
F= - \frac {1} {\beta} \ln Z \; ,
\end{equation}
where  $Z$ is the partition function
\begin{equation}
\label{eq45}
Z=Tr (e^{- \beta H}) \; .
\end{equation}
In the one-loop approximation we get :
\begin{equation}
\label{eq45a}
F_0= - \frac {1} {\beta} \ln Z_0, \;\;\;\; Z_0= Tr \left[e^{- \beta H_0}\right] \; .
\end{equation}
In our approximation the free energy is the sum of the photonic and fermionic 
contributions. Only the latter depends on the magnetic field. So that the calculation 
of the free energy reduces to consider the partition function of relativistic  
fermions in presence of the magnetic field.  \\
We consider the case $m>0$. A standard calculation gives :
\begin{equation}
\label{eq46}
Z_0= Tr  e^{-\beta(H_D+ \frac {V B^2} {2})} =  e^{-\beta \frac {V B^2} {2} }
(1+ e^{-\beta E_0})^{N_d} \; \prod_{n=1}^{\infty}  (1+ e^{- \beta E_n})^{2 N_d} 
\; e^{\beta E_n N_d} \;
\end{equation} 
where $N_d$ is the degeneracy of the Landau levels:
\begin{equation}
\label{eq47}
N_d= \frac {eB} {2 \pi} V \; .
\end{equation}
The free energy density  ${\cal {F}}_0(B) = \frac {F_0(B)} {V}$ is :
\begin{equation}
\label{eq48}
{\cal {F}}_0^{m>0} (B) = - \frac {eB} {\pi \beta} \sum_{n=1}^{\infty} \ln
(1+ e^{- \beta E_n}) - \frac {eB} {2 \pi \beta} 
\ln(1+ e^{- \beta E_0}) - \frac {eB} {2 \pi} \sum_{n=1}^{\infty}E_n 
+ \frac {B^2} {2} \; .
\end{equation}
In the same way we obtain :
\begin{equation}
\label{eq49}
{\cal {F}}_0^{m<0} (B) = - \frac {eB} {\pi \beta} \sum_{n=1}^{\infty} \ln
(1+ e^{- \beta E_n}) - \frac {eB} {2 \pi \beta} 
\ln(1+ e^{- \beta E_0}) - \frac {eB} {2 \pi} \sum_{n=0}^{\infty}E_n 
+ \frac {B^2} {2} \; .
\end{equation}
In Eqs.~(\ref{eq48}) and (\ref{eq49}) the first term on the right hand side
arises from the negative and positive Landau levels with $n>0$, the second one is the 
zero mode contribution, while the last two terms correspond to the zero temperature 
vacuum energy density. \\
To calculate the sums in Eqs.~(\ref{eq48}) and 
(\ref{eq49}) we expand the logarithm:
\begin{equation}
\label{eq50}
I_1(B)= - \frac {eB} {\pi \beta}  \sum_{n=1}^{\infty} \ln(1+ e^{-\beta E_n})
= - \frac { eB} {\pi \beta} \sum_{n=1}^{\infty} \sum_{k=1}^{\infty} 
\frac {(-1)^{k+1}} {k} e^{- \beta k E_n} \; ,
\end{equation}
and  use the integral representation~\cite{G-R} :

\begin{equation}
\label{eq51}
e^{- \sqrt{a b}} =  2 \sqrt{ \frac {a} {\pi}}  \int_0^{\infty} 
e^{-a x^2 - \frac {b} {x^2}} dx \;\;\;\;\; a>0, \;\;b>0  \; .
\end{equation}
This allows us to evaluate the sum over $n$:
\begin{equation}
\label{eq52}
\sum_{n=1}^{\infty} e^{- \beta k E_n} = \frac {1} {\sqrt{\pi}} 
\int_0^{\infty} e^{- \frac {x^2} {4} - \frac {1} {x^2} {\beta}^2 k^2 m^2} 
\frac {1} {e^{\frac {2 {\beta}^2 e B k^2} {x^2}} -1} \; .
\end{equation}
Therefore we have :
\begin{equation}
\label{eq53}
I_1(B) = - \frac {eB} { {\pi}^{ \frac {3} {2}}  \beta} \sum_{k=1}^{\infty}
\frac {(-1)^{k+1}} {k}  \int_0^{\infty} dx \;
e^{- \frac {x^2} {4} - \frac {1} {x^2} \beta^2 k^2 m^2} 
\frac {1}  {e^{ \frac {2 \beta^2 eB k^2} {x^2}} -1} \; .
\end{equation}
After subtracting  the contribution at $B=0$ we obtain :
\begin{eqnarray}
\label{eq54}
{\tilde {I}}_1(B) & = & I_1(B) - I_1(B=0) =  \nonumber \\
& = &
- \frac {eB} { {\pi}^{\frac {3} {2} }
\beta} \sum_{k=1}^{\infty} \frac {(-1)^{k+1}} {k} \int_0^{\infty} dx \;  
e^{ - \frac {x^2} {4} - \frac {1} {x^2} \beta^2 k^2 m^2}
\left[ \frac {1} { e^{ \frac {2 \beta^2 eB k^2} {x^2}} -1} - 
\frac {x^2} {2 eB \beta^2 k^2} \right] \; .
\end{eqnarray}
The sum over $k$ is rapidly convergent. So that this last expression is amenable to
 numerical evaluations. Putting it all together we finally get :
\begin{eqnarray}
\label{eq55}
{\tilde {\cal {F}}}_0^{m>0} (B) 
& = &
{\cal {F}}_0^{m>0} (B) -  {\cal {F}}_0^{m>0} (B=0) = 
\nonumber \\
& = & 
{\tilde {I}}_1(B) - \frac {eB} {2 \pi \beta} \ln (1+ e^{- \beta E_0})
+ {\tilde {\cal {E}}}_{m>0} (B) 
\end{eqnarray}
and
\begin{eqnarray}
\label{eq56}
{\tilde {\cal {F}}}_0^{m<0} (B) 
& = &
{\cal {F}}_0^{m<0} (B) -  {\cal {F}}_0^{m<0} (B=0) = 
\nonumber \\
& = & 
{\tilde {I}}_1(B) - \frac {eB} {2 \pi \beta} \ln (1+ e^{- \beta E_0})
+ {\tilde {\cal {E}}}_{m<0} (B) \; .
\end{eqnarray}
In Figures~4 and ~5 we display the free energy density, respectively Eq.~(\ref{eq55}) 
and Eq.~(\ref{eq56}), for different temperatures 
${\hat {T}} = \frac {T} {|m|}$ and $\alpha = \frac {|m|} {e^2}=0.1$, as a 
function of $\lambda = \frac {eB} {m^2}.$
\\
It is remarkable to observe that in the case $m<0$ by increasing the 
temperature the negative minimum does not disappear. This means that at high temperature
there is not restoration  of the symmetry broken at zero 
temperature. In other words, the spontaneous generation of the magnetic 
condensate survives even at infinite temperature.
Remarkably enough, the authors of Ref.~\cite{KANEMURA}, within a different
approach, found similar results also in presence of the Chern-Simon term . \\
It is interesting to study  the high temperature asymptotic expansion  
of the free energy density in the case of negative mass. 
Indeed we have :
\begin{equation}
\label{eq57}
I_1(B) = \frac {eB} {2 \pi \beta} \ln 2 + 
\frac {eB} {2 \pi} \sum_{n=1}^{\infty} E_n + {\cal {O}}(\beta^2) \; ;
\end{equation}
moreover
\begin{equation}
\label{eq58}
- \frac {eB} {2 \pi \beta} \ln(1+e^{-\beta E_0}) = - \frac {eB} {2 \pi \beta}
\ln 2 + \frac {eB} {4 \pi} |m| + {\cal {O}}(\beta^2)  \; .
\end{equation}
So that:
\begin{equation}
\label{eq59}
{\tilde {\cal {F}}}_0^{m<0} (B) = - 
\frac {eB} {4 \pi} |m| +
\frac {B^2} {2}  + {\cal {O}}(\beta^2) \; .
\end{equation}
We see that the linear term in $|m|$  gives rise to the negative minimum
at any finite temperature. Moreover, the slope of the linear 
term coincides with the one at  zero temperature.
This is clearly seen in  Fig.~6 where we display the free energy density
 without the classical energy $\frac {B^2} {2}$ for three different values of 
the temperature.  \\
The negative minimum of Eq.~(\ref{eq59}) is at
\begin{equation}
\label{eq60}
e B^* = \frac {e^2 |m|} {4 \pi} \; .
\end{equation}
As a consequence the negative condensation energy is :
\begin{equation}
\label{eq61}
{\tilde {\cal {F}}}_0^{m<0} (B^*) = - \frac {e B^*} {2} |m|= - \frac {e^2 |m|^2}
{32 \pi^2} \; .
\end{equation}
 The minimum of Eq.~(\ref {eq60}) coincides with the zero temperature minimum  
 in the "weak-coupling" region  $\alpha >>1$ :
\begin{equation}
\label{eq62}
\frac {e B^*} {m^2} = \frac {1} {4 \pi \alpha} \; .
\end{equation}
We would like to conclude this Section by stressing that our results show that the negative
minimum in the free energy density is due to the term linear in the magnetic field. This
term, in turn, is accounted for by the contributions due to the zero modes, and it is
present only for negative fermion mass. Moreover, we find that the thermal corrections
to the above mentioned coefficient are small and vanish at infinite temperature.
For these reasons we feel that  higher order  corrections do not should modify the 
spontaneous generation of the magnetic condensate .
\\
\\
\\
{\bf 3. P-Even Fermions }
\\
In three dimensions we have a realization of the 
Dirac algebra which is different from the $2 \times 2$ representation Eq.~(\ref{eq3}). 
In this representation the Dirac fermions are four-component spinor and the 
three $4 \times 4$ $\gamma$-matrices can be taken to be~\cite{Appelquist}:
\begin{equation}
\label{eq63}
\gamma^0= \left( \begin{array}{clcr}
                   \sigma_3  & \; \; \; 0 \\
                   0 & - \sigma_3  
                   \end{array} \right) \;\;\;
\gamma^1= \left( \begin{array}{clcr}
                    i \sigma_1  & \; \; \; 0 \\
                   0 & -i \sigma_1  
                   \end{array} \right) \;\;\;
\gamma^2= \left( \begin{array}{clcr}
                    i \sigma_2  & \; \; \; 0 \\
                   0 & -i \sigma_2  
                   \end{array} \right) \;\;\; .
\end{equation}                    
 The representation given by  Eq.~(\ref{eq63})
corresponds  to a reducible representation of  the Dirac algebra.
In this representation the fermionic mass term  $m {\bar {\psi}} \psi$ 
is parity conserving. Indeed,
comparing  Eq.~(\ref{eq3}) with Eq.~(\ref{eq63}) we see that a four-component fermion with
mass $m$ corresponds to two two-component fermions with mass $m$ amd -$m$ respectively. \\
As the first step, we need to solve the Dirac equation  Eq.~(\ref{eq8}) with
the Dirac algebra representation  Eq.~(\ref{eq63}).
Indeed, it is straightforward to obtain the orthonormal positive and negative energy 
solutions of Eq.~(\ref{eq8}). We get :

\begin{mathletters}
\begin{equation}
\label{eq64}
E= +E_n  \, , \; n \geq 1 \; \;  \psi^{(+)}_1 = \sqrt{\frac {E_n+m} {2 E_n}} 
 {e^{i p x_2 -\frac {1} {2} {\zeta}^2 }}
 \left( \begin{array}{c} 
N_n H_n(\zeta) \\
- \frac {\sqrt{E_n^2-m^2}} {E_n+m} N_{n-1} H_{n-1}(\zeta) \\
 0 \\ 0
\end{array} \right) 
\end{equation}
\begin{equation}
E= +E_n  \, , \; n \geq 1 \; \;  \psi^{(+)}_2 = \sqrt{\frac {E_n+m} {2 E_n}} 
 {e^{i p x_2 -\frac {1} {2} {\zeta}^2 }}
 \left( \begin{array}{c} 0 \\ 0 \\
- \frac {\sqrt{E_n^2-m^2}} {E_n+m} N_{n} H_{n}(\zeta) \\
N_{n-1} H_{n-1}(\zeta)
\end{array} \right) 
\end{equation}
\begin{equation}
E= -E_n  \, , \; n \geq 1 \; \;  \psi^{(-)}_1 = \sqrt{\frac {E_n+m} {2 E_n}} 
 {e^{-i p x_2 -\frac {1} {2} {\xi}^2 }}
 \left( \begin{array}{c} 
 \frac {\sqrt{E_n^2-m^2}} {E_n+m} N_{n} H_{n}(\xi) \\
N_{n-1} H_{n-1}(\xi)\\ 0 \\ 0 
\end{array} \right) 
\end{equation}
\begin{equation}
E= -E_n  \, , \; n \geq 1 \; \;  \psi^{(-)}_2 = \sqrt{\frac {E_n+m} {2 E_n}} 
 {e^{-i p x_2 -\frac {1} {2} {\xi}^2 }}
 \left( \begin{array}{c} 0 \\ 0 \\ 
N_n H_n(\xi) \\
 \frac {\sqrt{E_n^2-m^2}} {E_n+m} N_{n-1} H_{n-1}(\xi) 
\end{array} \right) 
\end{equation}
\end{mathletters}
with the same notations as in Eqs.~(\ref{eq10}), (\ref{eq11}) and $\xi=\sqrt{eB} 
\left(x_1+\frac {p} {eB} \right)$. The zero mode wave functions 
depend on sign of the mass term. If $m>0$ we have:

\begin{mathletters}
\begin{equation}
\label{eq65}
E_0=|m| \;\;\; \; \; \;
{\psi}_{0, m>0}^{(+)}= N_0 e^{ipx_2- \frac {1} {2} \zeta^2} 
\left( \begin{array}{c} 1 \\ 0 \\ 0 \\ 0 
\end{array} \right) 
\end{equation}
\begin{equation}
E_0=|m|  \;\;\; \; \; \;
{\psi}_{0, m>0}^{(-)}= N_0 e^{ipx_2- \frac {1} {2} \xi^2} 
\left( \begin{array}{c} 0 \\ 0 \\ 1 \\ 0 
\end{array} \right). 
\end{equation}
\end{mathletters}
For $m<0$ the zero mode wave functions are obtained by 
${\psi}_{0, m<0}^{\pm}= \gamma^5 {\psi}_{0, m>0}^{\pm}$, where:
\begin{equation}
\label{eq65bis}
\gamma^5= i \; \left( \begin{array}{clcr}
                  \; \; \; 0 \;  & \; \; I \\
                   -I &  \; \; 0  
                   \end{array} \right) \;\;\; .
\end{equation}
We stress that the 
density of the states for the zero modes is $\frac {|eB|} {2 \pi}$ 
while it is $\frac {|eB|} {\pi}$ for the modes with $n \geq 1$.
As already noticed, this is due to the fact that for $n \geq 1$ a four-component 
fermion corresponds to two independent degenerate two-component fermions.
\\
As in the previous section to calculate the condensate at $T=0$, we expand the 
field operator on the basis of the wave functions  Eqs.~(81a-d), (82a,b):
\begin{equation}
\label{eq66}
\psi= \int_{-\infty}^{+\infty} 
\frac {dp} {\sqrt{2 \pi}} [ a (0, p) {\psi}_{0,p}^{(+)} + 
b^{\dag}(0,p) {{\psi}_{0,p}^{(-)}}] + 
\sum_{n=1}^{\infty} \sum_{i=1,2} \int_{-\infty}^{+\infty} 
\frac {dp} {\sqrt{2 \pi}} 
[a_i(n,p) {{\psi}_{i,n,p}^{(+)}} + b_i^{\dag}(n,p) 
{{\psi}_{i,n,p}^{(-)}} ]
\end{equation}
where the operators $a_i$ and $b_i$ satisfy the anticommutation relations 
Eq.~(\ref{eq17}).  

It is possible to obtain the fermion condensate at $T=0$ through the  propagator 
$G(x,y)=\langle 0 | T \psi(x) {\bar {\psi}}(y)  | 0 \rangle$:
\begin{equation}
\label{eq67}
\langle 0 | {\bar {\psi}} \psi | 0 \rangle = - \lim_{x \rightarrow y} tr G(x,y)
\end{equation}
by using the Schwinger proper time approach~\cite{Schwinger}.
However, in the Furry's representation and in the one-loop approximation it is rather 
simple to obtain the condensate  at  $T=0$:
\begin{equation}
\label{eq68}
\langle 0 | {\bar {\psi}} \psi | 0 \rangle = - \frac {m} {|m|} \frac {eB} 
{2 \pi} - \frac {meB} {\pi} \sum_{n=1}^{\infty} \frac {1} {E_n} \; . 
\end{equation}
The massless limit of the condensate reads :
\begin{equation}
\label{eq69}
 \langle 0 | {\bar {\psi}} \psi | 0 \rangle |_{m \rightarrow 0} = 
- \frac {m} {|m|} \frac {eB} {2 \pi} \; .
\end{equation}
Although the massless limit of the fermion condensate is non-vanishing, it depends
on the sign of $m$. This means that the limit value does depend on the limiting
procedure. For instance, if one performs that symmetric massless limit, then one
gets a vanishing condensate.
As a matter of fact, we shall show below that at any finite temperature the
fermion condensate vanishes for $m \rightarrow 0$. 
To see that, we evaluate the thermal correction to the condensate. In our approximation 
and taking into account Eqs.~(\ref{eq39}), (\ref{eq40}) and (\ref{eq66}), 
it is straightforward to calculate the finite temperature fermion condensate. We get:
\begin{equation}
\label{eq70}
\langle {\bar {\psi}} \psi {\rangle}_{\beta} = - \frac {m} {|m|} \frac {eB} {2 
\pi} \tanh \left( \frac {\beta} {2} E_0 \right) - \frac {meB} {\pi} 
\sum_{n=1}^{\infty} \frac {1} {E_n} \tanh \left( \frac {\beta} {2} E_n \right).
\end{equation}
This expression is in agreement with Ref.~\cite{D-H}, where the authors calculate the 
value  of the condensate at finite temperature using the thermo field 
dynamics~\cite{U-M-T} approach with  a non-vanishing chemical potential. \\
Let us consider the limits $m \rightarrow 0$ and $\beta \rightarrow \infty$.
We have:
\begin{equation}
\label{eq71}
\lim_{m \rightarrow 0} \lim_{ \beta \rightarrow \infty}
\langle {\bar {\psi}} \psi \rangle_{\beta} = 
- \frac {m} {|m|} \frac {eB} {2 \pi} \; ,
\end{equation}
which, of course, agrees with Eq.~(\ref{eq69}). On the other hand, for any temperature
we have:
\begin{equation}
\label{eq72}
\left. \;\;\;\;\; \right.
\lim_{m \rightarrow 0} \langle {\bar {\psi}} \psi \rangle_{\beta}= 0 \; .
\end{equation}
Thus, we see that the ($2+1$)-dimensional P-even massless fermion 
condensate in presence of a constant magnetic field is highly unstable. In fact 
the thermal condensate  disappears as soon as a heat bath is introduced. \\ 
Let us, now, turn on the effective potential. Again we need to
evaluate the Dirac energy $E_D$. To this end we rewrite the one-loop
Hamiltonian Eq.~(\ref{eq6b}) by using the expansion Eq.~(\ref{eq66}).
For the Dirac Hamiltonian $H_D$ we find :
\begin{eqnarray}
\label{eq72bis}
H_D
& = &
\int dp \, E_0 \, [a^{\dag}(0,p) a(0,p) + b^{\dag} (0,p) b(0,p) ] + \sum_{n=1}^{\infty} 
\sum_{i=1,2} \int dp \,  E_n \, [a^{\dag}_i(n,p) a_i(n,p) + b^{\dag}_i(n,p) b_i(n,p) ] 
\nonumber \\
&  & 
- \frac {eB} {4 \pi} V \; \; [E_0
+ 2 \sum_ {n=1}^{\infty}E_n] \; .
\end{eqnarray}
As a consequence the Dirac energy defined by Eq.~(\ref{eq25}) is:
\begin{equation}
\label{eq72tris}
E_D(B)=- \frac {eB} {2 \pi}\frac {V} {2}[E_0
+ 2 \sum_ {n=1}^{\infty}E_n]
\end{equation}
Proceeding as in the previous Section we define the vacuum energy density. We get: 
\begin{equation}
\label{eq72poker}
{\tilde {\cal {E}}}(B) = {\cal {E}}(B) - {\cal {E}}(0)= \frac {B^2} 
{2} + \frac {(eB)^{\frac {3} {2}}} {2 \pi} 
g\left(\frac {eB} {m^2}\right) + \frac {eB} {4 \pi} |m| \; ,
\end{equation}
where the function $g(\lambda)$ is given by Eq.~(\ref{eq32}). \\
Note that, if we rewrite the last term on the right hand size of Eq.~(\ref{eq72poker})
by using the integral representation  Eq.~(\ref{eq32}), and after some manipulations, we
get:
\begin{equation}
\label{eq72five}
{\tilde {\cal {E}}}(B) =  \frac {B^2} {2}  +  \frac {1} {8 \pi^{\frac {3} {2}} } \;
\int_0^{\infty} \frac {ds} 
{s^{\frac {5} {2}} } \; e^{- m^2 s} \; 
\left[ (eBs) \coth(eBs) - 1 \right] \; .
\end{equation}
Equation~(\ref{eq72five}) agrees with the calculation by Redlich~\cite{Redlich}. \\
As concern the finite temperature effective potential 
we follow the 
same steps as in the previous Section. The partition function is :
\begin{equation}
\label{eq72ter}
Z_0= Tr  e^{-\beta(H_D+ \frac {V B^2} {2})} =  e^{- \beta \frac {B^2} {2} V} \,
(1- e^{-\beta E_0})^{d} \; e^{\beta \frac {d} {4} E_0 }\; 
\prod_{n=1}^{\infty}  (1- e^{- \beta E_n})^{4 d} 
\;   e^{\beta \frac {d} {2}  E_n }, 
\end{equation} 
where $d= \frac {eB} {\pi}$. So that the free energy density is: 
\begin{equation}
\label{eq72quater}
{\cal {F}}_0 (B) = - \frac {4 d} {\beta} \sum_{n=1}^{\infty} \ln
(1+ e^{- \beta E_n}) - \frac {d} {\beta} 
\ln(1+ e^{- \beta E_0}) - \frac {d} {4} E_0 -\frac {d} {2} \sum_{n=1}^{\infty}E_n 
+ \frac {B^2} {2} \; .
\end{equation}
Subtracting the contribution at $B=0$, we obtain 
\begin{eqnarray}
\label{eq73}
{\tilde {\cal {F}}}_0 (B) 
& = &
{\cal {F}}_0(B) -  {\cal {F}}_0 (B=0) = 
\nonumber \\
& = & 
 4 {\tilde {I}}_1 (B) - \frac {eB} { \pi \beta}
\ln (1+ e^{- \beta m}) + \frac { (eB)^{\frac {3} {2}}} {2 \pi} 
g\left( \frac {eB} 
{m^2} \right) + \frac {eB} {4 \pi} |m| + \frac {B^2} {2}
\end{eqnarray}
where ${\tilde {I}}_1 (B)$ has been already defined in the previous Section. 
Note that, as expected, the zero-temperature limit of Eq.~(\ref{eq73}) reduces to
Eq.~(\ref{eq72poker}). \\
In Figure 7 we plot the free energy density as a function of $\lambda$ for three 
values of temperature ${\hat {T}}$.
As it is evident the spontaneous generation of an uniform magnetic condensate is
not energetically favorable neither at zero temperature nor at any finite temperature.
\\
\\
\\
{\bf 4. Dynamical Generation of Mass in External Magnetic Field  }
\\
In this Section we investigate the spontaneous generation of the fermion mass in
presence of an external magnetic field. Let us consider massless Dirac fermions
in an external constant magnetic field. The relevant Lagrangian is now :

\begin{equation}
\label{eq79}
{\cal{L}}_{QED} (x) = \bar {\psi} (x) \gamma^{\mu} (i \partial_{\mu} - e A_{\mu})
\psi (x) - \frac {1}{4} F_{\mu \nu} (x)  F^{\mu \nu} (x)
\end{equation}
Recently there has been renewed interest in the dynamical symmetry breaking in 
$(2+1)$-dimensional {\it QED}. In particular it turns
out that in planar{\it QED}  a constant magnetic field leads to the dynamical 
generation of a P-even mass for four-component Dirac 
fermions~\cite{FARAKOS}. \\
In order to investigate the dynamical symmetry breaking we must consider 
the effective action for composite operators~\cite{C-J-T}. For the problem
of spontaneous generation of a fermion mass the effective action will depend 
on the complete fermion propagator of the theory. However we restrict ourselves 
to the case of a constant external source coupled to the local operator 
$\bar { \psi (x)} \psi (x)$. \\
In this case the external source acts like a fermion mass $m$, so that the 
effective action depends on the full propagator for massive fermion fields.
Moreover, owing to the translation invariance of the problem, we need to 
consider the effective potential. According to Ref.~\cite{C-J-T},
the relevant quantity turns out to be: 
\begin{equation}
\label{eq80}
\Omega(B,m)=V_{\mathrm eff}(B,m)-V_{\mathrm eff}(B,0) \; ,
\end{equation}
where $V_{\mathrm eff}(B,m)$ is the effective potential in presence of both the magnetic
field and the constant fermion mass. 
Moreover we have the well known physical interpretation that:
\begin{equation}
\label{eq81}
V_{\mathrm eff}(B,m)= \frac {1} {V} \frac {\langle 0_m | H | 0_m \rangle } 
{\langle 0_m | 0_m \rangle}
\end{equation}
where $| 0_m \rangle$ indicates the ground state of the massive theory, and $H$
is the massless Hamiltonian: 
\begin{equation}
\label{eq82}
H=\int d^2 x  \left[\frac {B^2} {2} + \psi^{\dag} (-i {\vec {\alpha}} 
\cdot \vec {\nabla}- e {\vec {\alpha}} \cdot \vec {A}) \psi \right] \; .
\end{equation}
Our strategy is to consider a trial vacuum state which depends on the 
variational parameter $m$, and thus determine the effective potential by 
minimizing the vacuum energy density.
\\
We are interested in the case of a given uniform external magnetic field. In 
this case, in the one-loop approximation the Hamiltonian reduces to:
\begin{equation}
\label{eq83}
H=\int d^2 x  \left[\frac {B^2} {2} + \psi^{\dag} (-i {\vec {\alpha}}
\cdot \vec {\nabla}- e {\vec {\alpha}} \cdot {\vec {{\bar A}}})\psi \right]
\end{equation}
where ${\vec {{\bar A}}}$ is given by Eq.~(\ref{eq6}).
To evaluate the effective potential we need to calculate the expectation value 
of the massless Hamiltonian Eq.~(\ref{eq83}) on the massive fermion ground 
state.  \\
The case of P-even mass term has been extensively discussed in the 
literature~\cite{FARAKOS}. Therefore we restrict to the case of P-odd
Dirac fermions. \\
Let us consider, firstly, the case of a negative mass term $m=-|m|$. 
The massive ground 
state $|0_m \rangle$ have been already constructed in Sect.~2. Using the 
results of that Section we find: 
\begin{equation}
\label{eq84}
\frac {1} {V}  
\frac {\langle 0_m | H| 0_m \rangle }
{\langle 0_m | 0_m \rangle } = \frac {B^2} {2} - \frac {e B}
{2 \pi} \sum_{n=0}^{\infty} \sqrt{2 n e B +m^2} + \frac {|m|} {V}  
\frac {\langle 0_m | {\bar {\psi}}(\vec {x}) \psi (\vec {x}) | 0_m \rangle}
{\langle 0_m | 0_m \rangle }
\end{equation}
From Eq. (\ref{eq21}) it follows:
\begin{equation}
\label{eq85}
\frac {1} {V} 
\frac {\langle 0_m | H| 0_m \rangle }
{\langle 0_m | 0_m \rangle } = \frac {B^2} {2} - \frac {e B}
{2 \pi} \sum_{n=1} \sqrt{2 n e B +m^2} +
m^2 \frac {eB} {2 \pi} \sum_{n=1}^{\infty} \frac {1}
{\sqrt{2neB +m^2}}
\end{equation}
It easy to check that the same result follows in the case of positive mass 
term. So that we get
\begin{equation}
\label{eq86}
V_{\mathrm eff}(B,m)= \frac {B^2} {2} - \frac {eB} {2 \pi} \sum_{n=1}^{\infty} 
\sqrt{2neB+m^2} + 
\frac {m^2 eB} {2 \pi} \sum_{n=1}^{\infty}
\frac {1} {\sqrt{2neB+m^2}}.
\end{equation}
Observing that 
\begin{equation}
\label{eq87}
V_{\mathrm eff}(B,0)= \frac {B^2} {2} + \frac {\sqrt{2}} {8 \pi^2} 
\zeta (\frac {3} {2}) (eB)^{\frac {3} {2}}
\end{equation}
we get finally
\begin{equation}
\label{eq88}
\Omega(B,m)= - \frac {eB} {2 \pi} \sum_{n=1}^{\infty} 
\sqrt{2neB+m^2}+
\frac {m^2 eB} {2 \pi} \sum_{n=1}^{\infty}
\frac {1} {\sqrt{2neB+m^2}}
- \frac {\sqrt{2} \zeta (\frac {3} {2})} {8 \pi^2} (eB)^{\frac {3} {2}}.
\end{equation}
Note that, as expected, the effective potential $\Omega(B,m)$
depends on $|m|$. By using  the integral representation Eq.~(\ref{eq29})
together with
\begin{equation}
\label{eq89}
\frac {1} {\sqrt{a}} = \int_0^{\infty} \frac {ds} {\sqrt{\pi s}}e^{-as},
\end{equation}
we obtain the following integral representation for the effective
potential:
\begin{equation}
\label{eq90}
\Omega(B,m) = |m|^3 \left[ J(\lambda) - \frac {\sqrt{2}} 
{8 \pi^2} \zeta(\frac {3} {2}) \lambda^{\frac {3} {2}}\right]
\end{equation}
where we recall that $\lambda = \frac {eB} {m^2}$ and 
\begin{equation}
\label{eq91}
J(\lambda)= - \frac {\lambda^2} {\pi} \int_0^{\infty} \frac {d y} 
{\sqrt{ \pi y}} e^{y} \left[\frac {e^{-2 \lambda y}} 
{(1-e^{-2 \lambda y})^2} - \frac {1} {(2 \lambda y)^2} \right].
\end{equation}
In order to obtain the small $|m|$ behaviour of the effective potential, 
we note that
\begin{equation}
\label{eq92}
J(\lambda) \stackrel{\lambda \rightarrow  \infty}{\sim} 
\frac {\sqrt{2}} {8 \pi^2} \lambda^{\frac {3} {2}} 
\zeta(\frac {3} {2}) + \frac {\sqrt{2}} {8 \pi} \lambda \zeta(\frac {1} {2})
+ {\cal {O}} (\frac {1} {\sqrt {\lambda}}).
\end{equation}
So that for small $|m|$ we get 
\begin{equation}
\label{eq93}
\Omega(B,m)= \frac {\sqrt{2}} {8 \pi} \zeta(\frac {1} {2}) \sqrt{eB}m^2+
{\cal {O}} (m^4).
\end{equation}
On the other hand for large $|m|$ we find
\begin{equation}
\label{eq94}
J(\lambda) \stackrel{\lambda \rightarrow  0}{\sim}  
\frac {\lambda^2} {12 \pi}
+ \frac {\lambda^3} {6 \pi} + {\cal {O}} \left(\lambda^4 \right),
\end{equation}
so that
\begin{equation}
\label{eq95}
\Omega(B,m)= - \frac {\sqrt {2}} {8 \pi^2} \zeta(\frac {3} {2}) 
(eB)^{\frac {3} {2}} + {\cal {O}} (\frac {1} {|m|}).
\end{equation}
Equation (\ref {eq93}) tell us that the one-loop effective potential 
$\Omega(B,m)$ displays an absolute maximum at $m=0$. Moreover it turns out that 
$\Omega(B,m)<0$ (see Fig. 8). So that it is energetically 
fovourable to generate a constant fermion mass.  \\
Remarkably, a straightforward calculation shows that in the one-loop 
approximation the effective potential $\Omega(B,m)$ given by Eq. (\ref {eq88})
accounts for the dynamical generation of a constant mass 
also in the case of four 
component fermions. As a consequence the remarkable phenomenon of the dynamical 
generation of a P-even fermion mass in presence of an external constant 
magnetic fields displays itself even in the one-loop approximation. 
Note that the minimum of the one-loop effective potential is obtained for 
infinite value of the mass. However, we feel that this result is an artefact of 
the one-loop approximation. In fact, including the higher-order corrections 
should give rise to a negative minimum for finite value of the fermion mass. 
It is worthwhile to briefly compare our approach with the papers
in Ref~\cite{FARAKOS}. The papers in Ref~\cite{FARAKOS} deal with 
four-component P-even Dirac fermions coupled with $(2+1)$-dimensional
electromagnetic fields. In these papers it is studied the full
Schwinger-Dyson equations in the limit of strong magnetic field where
the lowest Landau level accounts for the fermion
dynamics. On the other hand, in our approach we restrict to the one-loop 
approximation, but we take into account all Landau levels, reaching conclusion 
similar to the ones in Ref~\cite{FARAKOS}. Thus we feel that our study 
corroborates the validity of the approximation of Ref~\cite{FARAKOS}. 
\\
\\
\\
{\bf 5. Conclusions }
\\
In this paper we have discussed in details the dynamics of $(2+1)$-dimensional
Dirac fermions coupled to an external magnetic field in the one-loop
approximation both at zero and finite temperature. \\
In particular we investigated in the two 
formulations the spontaneous generation af the magnetic condensate both at $T=0$ and 
$T\not= 0$.
Our results show that in the one-loop approximation the remarkable phenomenon of spontaneous
generation of an uniform magnetic condensate takes place only in the
case of Dirac fermions with negative P-odd mass. Moreover the magnetic
condensation survives the thermal corrections even at infinite temperature.
 In addition  at high 
temperature the thermal fluctuations tend to increase the condensation energy 
and the strength of the induced magnetic  field . \\
 It is worthwhile to note that the effective system composed of half
 filled zero modes at zero temperature accounts for the high temperature
 limit of the free energy. So that we see that the dynamics of the zero
 modes is responsible for the dynamical magnetic condensation both at 
 zero and high temperatures.
Remarkably, as discussed in the Introduction, it turns out that  P-odd 
Dirac fermions are relevant for
the descriptions of zero modes localized on a domain wall in the case
of four dimensional fermions coupled with a scalar field.
Indeed it has been suggested~\cite{CEA5} that the spontaneous generation of an uniform
magnetic condensate gives rise to  ferromagnetic domains walls at the
electroweak phase transition . This, in turns, may have far
reaching consequences during the electroweak phase transition in the early
universe. \\
We also showed that even in the one-loop approximation a constant magnetic
field is a strong catalyst for the spontaneous generation of a fermion mass.
In particular we find that the spontaneous generation of a constant fermion
mass does not depend on the spinorial representation for the Dirac matrices.
So that an external constant magnetic field induces the
generation of both a constant P-even and P-odd fermion mass. \\
Finally, we discussed the massless limit of the fermion condensate for both 
P-even and P-odd Dirac fermions in an external magnetic field. For both
formulations we find that the massless limit of the condensate is 
non-analytic. Remarkably, it turns out that in presence of a heat bath,
even with vanishing temperature, the fermion condensate survives to the
massless limit only for P-odd two-component fermion fields. Moreover, in
that case the massless limit is symmetric. \\
In conclusions, our results show that, within the one-loop approximation,
the unique theory displaying a non-trivial ground state turns out to be
the three-dimensional quantum electrodynamics in interaction with 
two-component Dirac fermion fields.
\newcommand{\InsertFigure}[2]{\newpage\phantom{.}\vspace*{-3cm}%
\phantom{.} \vspace*{5.truecm}
\vspace*{-8.truecm}
\begin{center}\mbox{%
\epsfig{bbllx=5.truecm,bblly=7.8truecm,bburx=16.5truecm,bbury=28.truecm,%
height=18.truecm,figure=#1}}
\end{center}\vspace*{-.1truecm}%
\parbox[t]{\hsize}{\small\baselineskip=0.5truecm\hskip0.5truecm #2}}

\InsertFigure{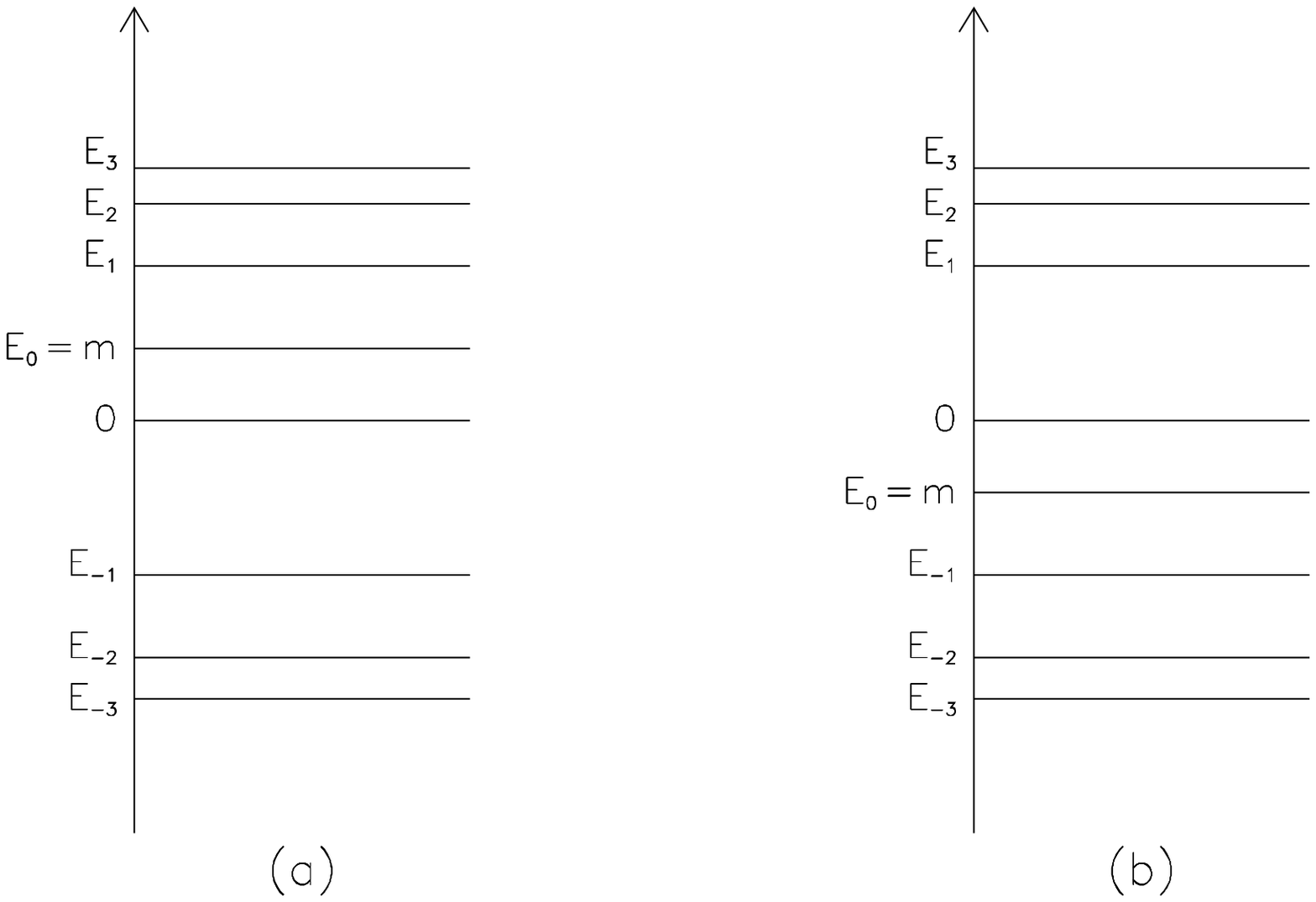}%
{FIG.~1           The Landau levels for Dirac fermions in a constant magnetic 
                  field: $a) \; m>0$ and
                  $b) \; m<0$.}   

\newcommand{\InsertFigureBis}[2]{\newpage\phantom{.}\vspace*{-2cm}%
\phantom{.} \vspace*{6.truecm}
\hspace*{.3truecm} ${{\tilde {\cal {E}}}}_{m<0} $
\vspace*{-10.truecm}
\begin{center}\mbox{%
\epsfig{bbllx=2.truecm,bblly=4.truecm,bburx=16.5truecm,bbury=28.truecm,%
height=18.truecm,figure=#1}}
\end{center}\vspace*{+.1truecm}%
\parbox[t]{\hsize}{\small\baselineskip=0.5truecm\hskip0.5truecm #2}}

\InsertFigureBis{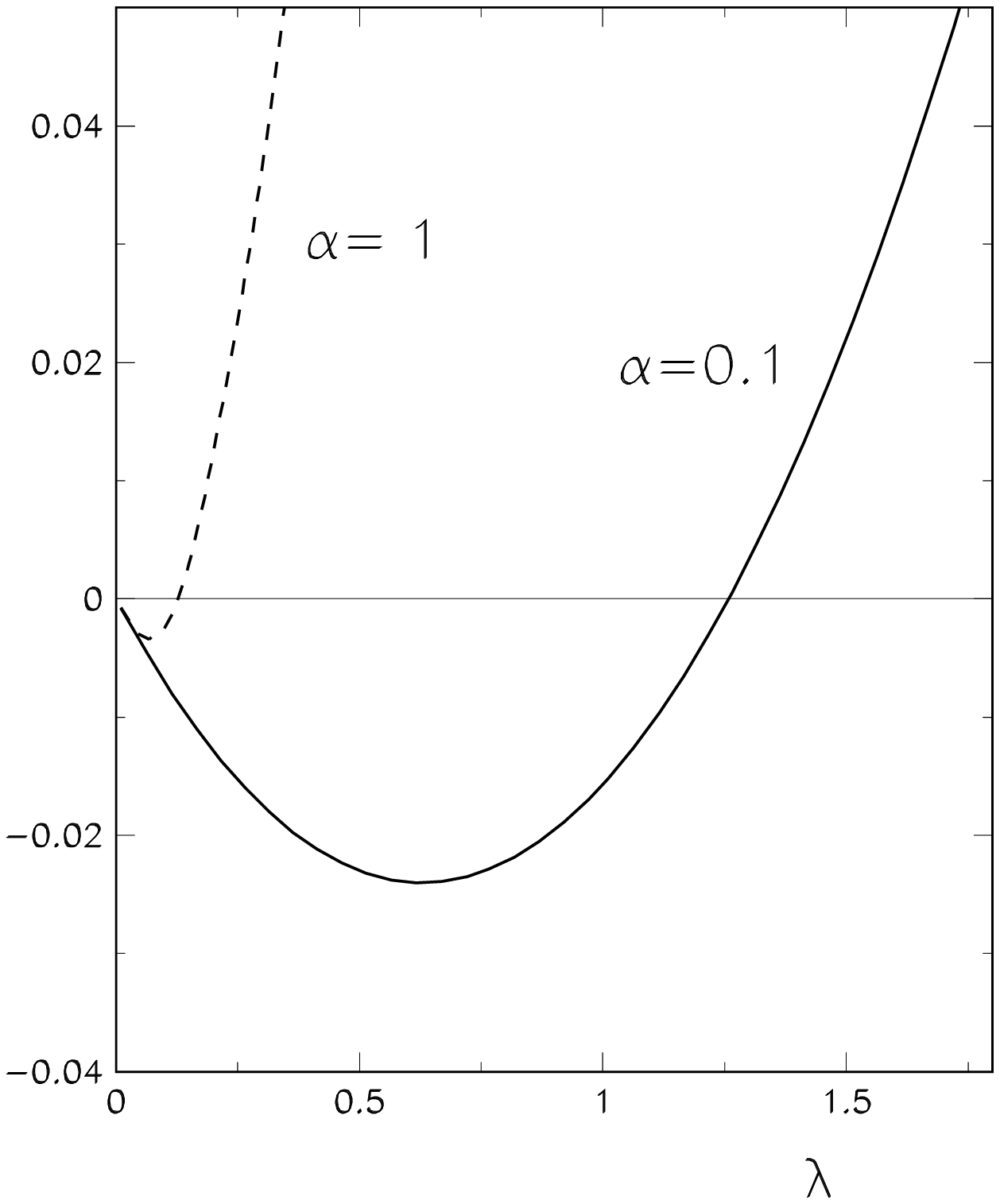}%
{FIG.~2           Plot of the energy density  
                  $\frac {{{\tilde {\cal {E}}}}_{m<0}} {|m|^3}$ as a function 
                  of $\lambda$ for two values of $\alpha= \frac {|m|} {e^2}$:
                  full line 
                  $\alpha=0.1$, dashed line $\alpha=1$.}
\newcommand{\InsertFigureTer}[2]{\newpage\phantom{.}\vspace*{-2cm}%
\phantom{.} \vspace*{6.truecm}
\hspace*{-.3truecm} ${{\tilde {\cal {E}}}}_{m>0} $
\vspace*{-6.truecm}
\begin{center}\mbox{%
\epsfig{bbllx=6.truecm,bblly=6.truecm,bburx=17.5truecm,bbury=28.truecm,%
height=18.truecm,figure=#1}}
\end{center}\vspace*{+.1truecm}%
\parbox[t]{\hsize}{\small\baselineskip=0.5truecm\hskip0.5truecm #2}}

\InsertFigureTer{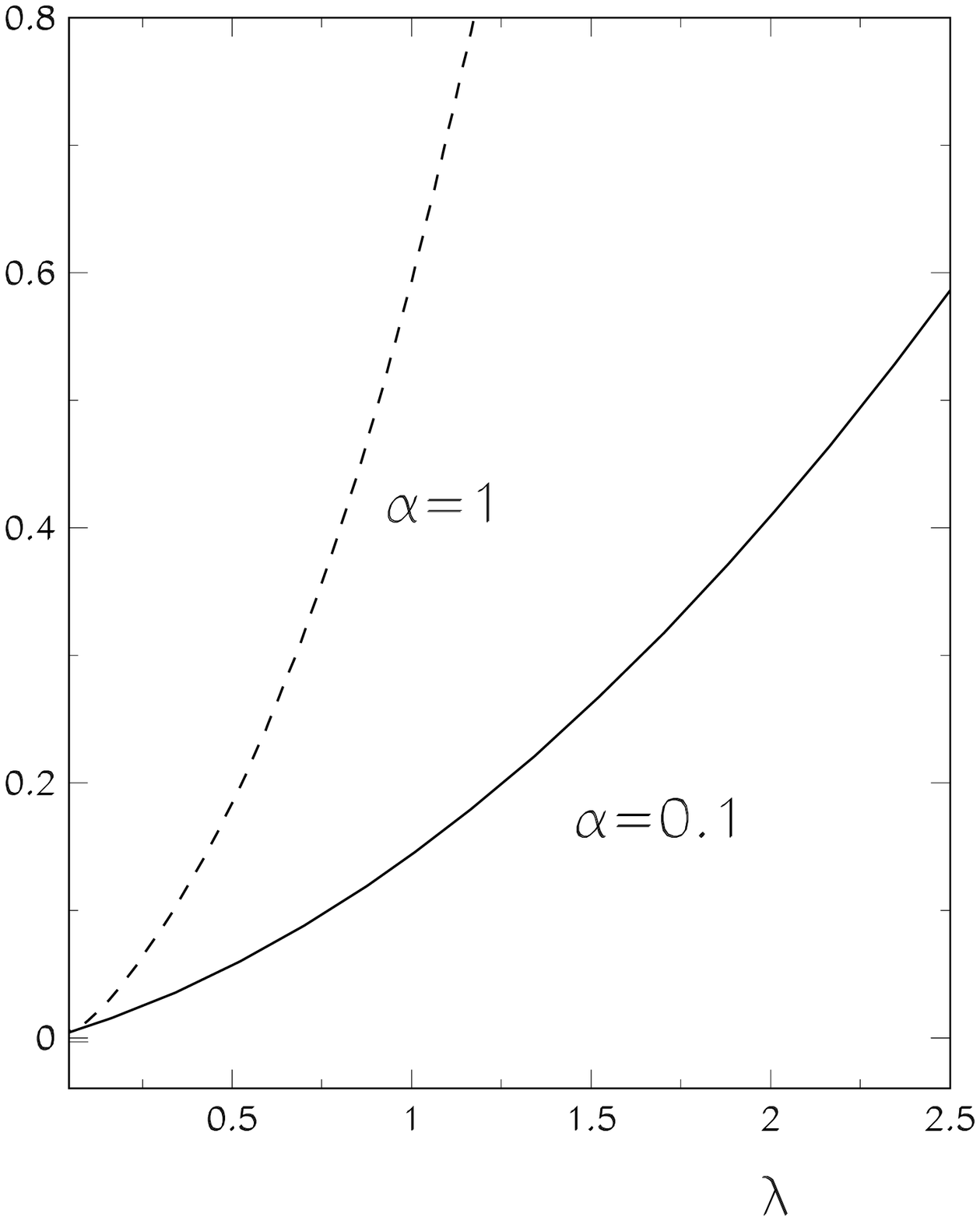}%
{FIG.~3           Plot of the energy density  
                  $\frac {{{\tilde {\cal {E}}}}_{m>0}} {|m|^3}$ as a function 
                  of $\lambda$ for two values of $\alpha= \frac {|m|} {e^2}$:
                  full line $\alpha=0.1$, dashed line $\alpha=1$.}
\newcommand{\InsertFigureB}[2]{\newpage\phantom{.}\vspace*{-2cm}%
\phantom{.} \vspace*{6.truecm}
\hspace*{.3truecm}  ${\tilde {{\cal {F}}}}_0^{m>0}$
\vspace*{-7.truecm}
\begin{center}\mbox{%
\epsfig{bbllx=4.truecm,bblly=5.6truecm,bburx=19.truecm,bbury=29.9truecm,%
height=18.truecm,figure=#1}}
\end{center}\vspace*{+.1truecm}%
\parbox[t]{\hsize}{\small\baselineskip=0.5truecm\hskip0.5truecm #2}}

\InsertFigureB{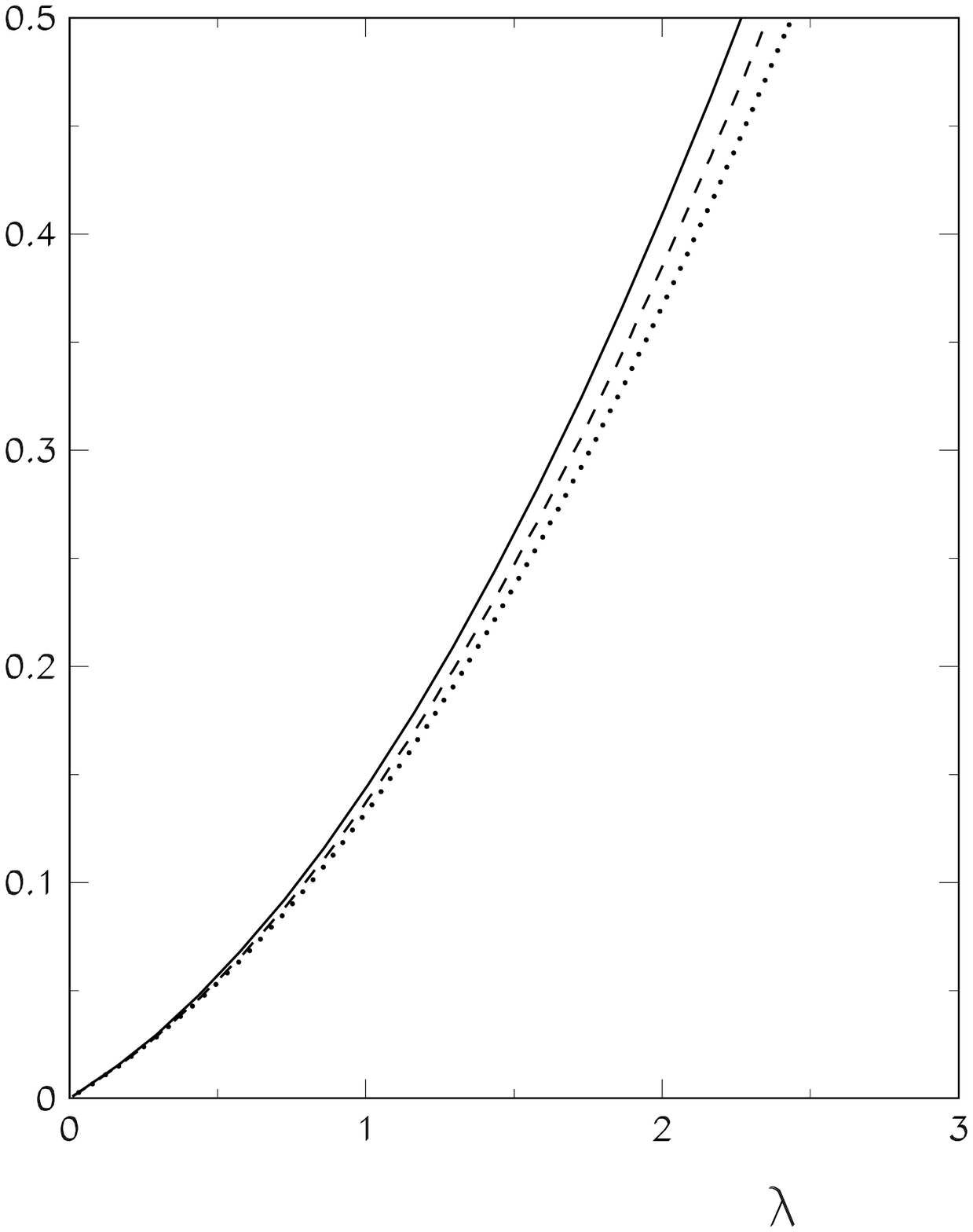}%
{FIG.~4           Free energy density (in units of $|m|^3$) versus $\lambda$
                  for $\alpha=0.1$ and $m>0$. Full line $T=0$, dashed line 
                  ${\hat {T}} =1$,  dotted line ${\hat {T}} =5$.}

\newcommand{\InsertFigureA}[2]{\newpage\phantom{.}\vspace*{-3cm}%
\phantom{.} \vspace*{5.truecm}
\hspace*{.3truecm} ${\tilde {{\cal {F}}}}_0^{m<0} $ 
\vspace*{-9.5truecm}
\begin{center}\mbox{%
\epsfig{bbllx=2.truecm,bblly=2.5truecm,bburx=18.truecm,bbury=27.5truecm,%
height=18.truecm,figure=#1}}
\end{center}\vspace*{-.1truecm}%
\parbox[t]{\hsize}{\small\baselineskip=0.5truecm\hskip0.5truecm #2}}

\InsertFigureA{a10_fig5.ps}%
{FIG.~5    Free energy density (in units of $|m|^3$) versus $\lambda$ for
           $\alpha=0.1$ and $m<0$. 
           \par
          Full line $T=0$, dashed line 
           ${\hat {T}}=1$, dotted line ${\hat {T}}=5$ and dashed-dotted line 
           ${\hat {T}}\rightarrow \infty$.}
\newcommand{\InsertFigureC}[2]{\newpage\phantom{.}\vspace*{-3cm}%
\phantom{.} \vspace*{5.truecm}
\hspace*{.2truecm} 
\hspace*{.3truecm} ${\tilde {{\cal {F}}}}_0^{m<0}$ 
\vspace*{-9.5truecm}
\begin{center}\mbox{%
\epsfig{bbllx=1.5truecm,bblly=2.8truecm,bburx=17.5truecm,bbury=28.truecm,%
height=18.truecm,figure=#1}}
\end{center}\vspace*{-.1truecm}%
\parbox[t]{\hsize}{\small\baselineskip=0.5truecm\hskip0.5truecm #2}}

\InsertFigureC{a10_fig6.ps}%
{FIG.~6              Free energy density (in unit of $|m|^3$) versus
		     $\lambda$ for $\alpha=0$ and $m<0$. Temperature values as in
		     FIG. 5.}
\newcommand{\InsertFigureD}[2]{\newpage\phantom{.}\vspace*{-2cm}%
\phantom{.} \vspace*{6.truecm}
\hspace*{.2truecm}  ${\tilde {{\cal {F}}}}_0$
\vspace*{-8.truecm}
\begin{center}\mbox{%
\epsfig{bbllx=4.truecm,bblly=7.truecm,bburx=19.truecm,bbury=31.truecm,%
height=18.truecm,figure=#1}}
\end{center}\vspace*{+.1truecm}%
\parbox[t]{\hsize}{\small\baselineskip=0.5truecm\hskip0.5truecm #2}}

\InsertFigureD{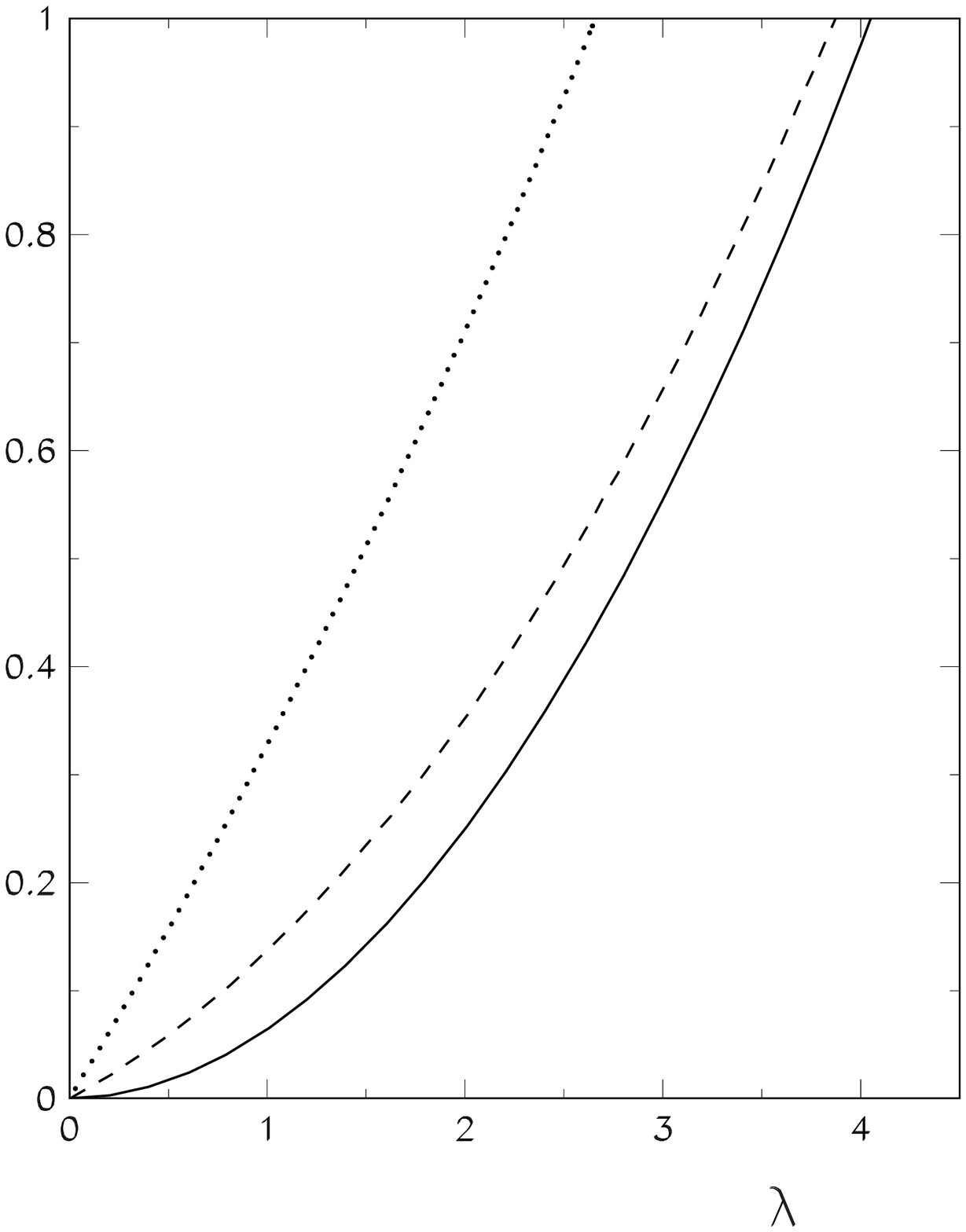}%
{FIG.~7           Free energy density in the P-even formulation 
                  (in units of $|m|^3$) versus $\lambda$
                  for $\alpha=0.1$. Full line $T=0$, dashed line 
                  ${\hat {T}} =1$,  dotted line ${\hat {T}} =2$.}
\newcommand{\InsertFigureE}[2]{\newpage\phantom{.}\vspace*{-2cm}%
\phantom{.} \vspace*{6.truecm}
\vspace*{-8.truecm}
\begin{center}\mbox{%
\epsfig{bbllx=4.truecm,bblly=7.truecm,bburx=19.truecm,bbury=31.truecm,%
height=18.truecm,figure=#1}}
\end{center}\vspace*{+.1truecm}%
\parbox[t]{\hsize}{\small\baselineskip=0.5truecm\hskip0.5truecm #2}}

\InsertFigureE{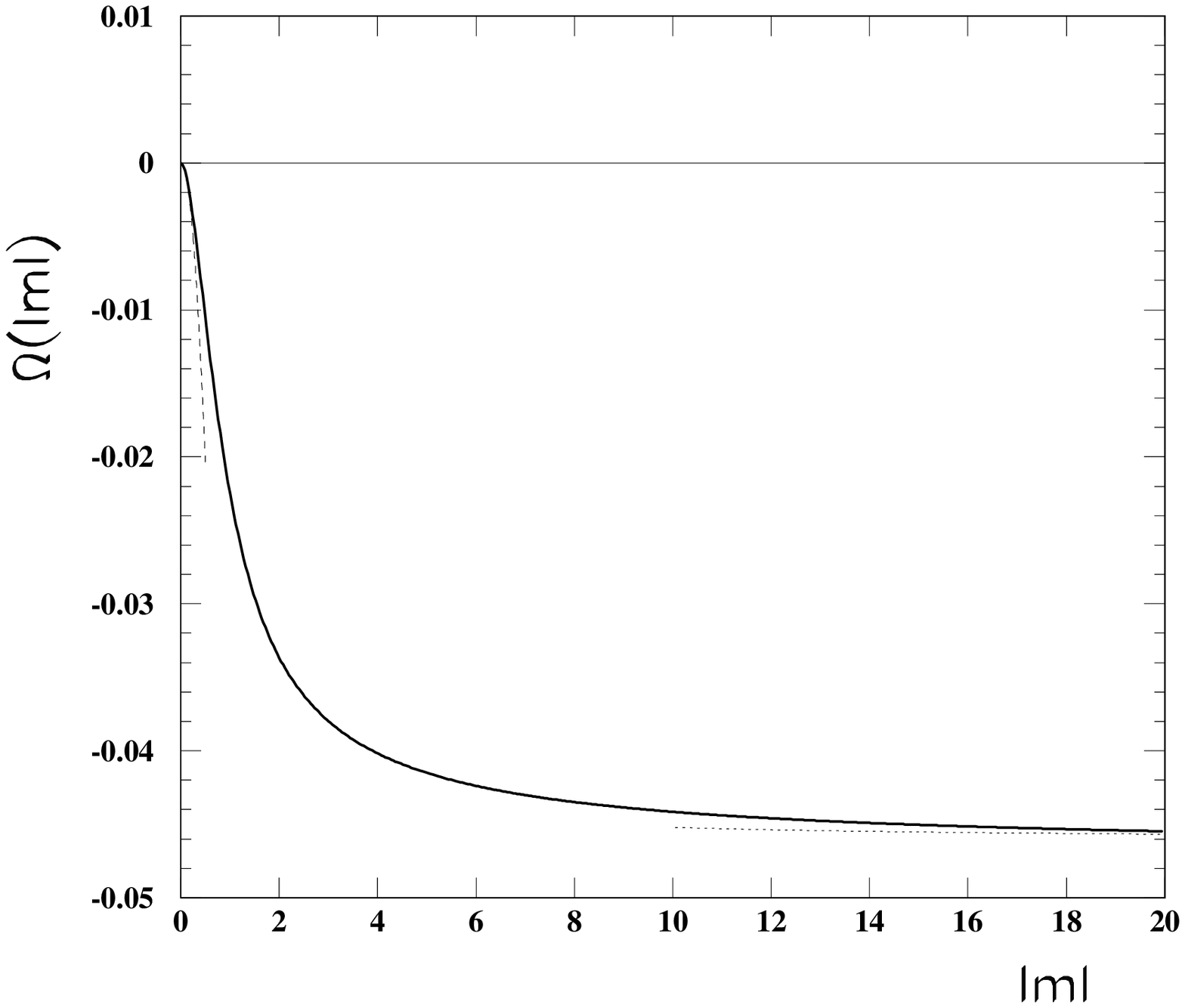}%
{FIG.~8            The effective potential $\Omega(B,m)$  
                   versus $|m|$
                  for $eB=1$. Dashed and dotted lines are the small and large $|m|$
                   expansions respectively.}

\end{document}